\documentstyle[epsf,12pt]{article}

\begin{document}

\begin{titlepage}
\begin{flushright}
{\large \bf UCL-IPT-97-10}
\end{flushright}
\vskip 2cm
\begin{center}

{\Large \bf Effects of the magnetic dipole moment of charged vector 
mesons in their radiative decay distribution}
 \vskip 1cm

{\large G. L\'{o}pez Castro$^{a,b}$ and G. Toledo S\'anchez$^b$} \\

$^a$ {\em Institut de Physique Th\'eorique, Universit\'e catholique de
Louvain,}\\ {\em  B-1348 Louvain-la-Neuve, Belgium} \\

$^b$ {\em Departamento de F\'\i sica, Centro de Investigaci\'on y de  
Estudios} \\ {\em Avanzados del IPN, Apdo. Postal 14-740, 07000 M\'exico, 
D.F., M\'exico}

\end{center}

\vskip 2cm

\begin{abstract}
We consider the effects of anomalous magnetic dipole moments of vector 
mesons in the decay distribution of photons emitted in 
two-pseudoscalar decays of charged vector mesons. By choosing a 
kinematical configuration appropriate to isolate these effects 
from model-dependent and dominant bremsstrahlung contributions, 
we show that this method can provide a valid alternative for a 
measurement of the unknown magnetic dipole moments of charged vector 
mesons.
\end{abstract}


\end{titlepage}%

\medskip

Electromagnetic multipole moments are important static properties that 
characterize particles and nuclei. While the electromagnetic current 
conservation imposes that the total electric charge must be conserved in 
a given reaction, its higher multipoles are not fixed in general 
by theoretical requirements and must be determined from experimental 
measurements. For elementary particles, the magnetic dipole moments of 
$e^-$, $\mu^-$ are measured with high precision 
\cite{pdg} whereas better constrains on the magnetic dipole and electric 
quadrupole moments of $W^{\pm}$ gauge bosons are becoming available from 
LEP2 and Tevatron colliders. 
In the case of hadrons, only the magnetic dipole moments of quasi-stable 
baryons have been measured, while those of hadronic resonances
 remain unknown \cite{pdg}.

 The spin precession technique \cite{bmt} used to measure the magnetic 
moments of octet baryons and the $\Omega^-$, is not applicable in the case 
of hadronic resonances due to their very short lifetimes ( $\leq 
10^{-18}$ sec). An alternative method based on photon emission off 
hadrons \cite{kpz} can be used in the later case, because the 
photon carries information on higher multipoles of emitting 
particles\footnote{In fact, this method is used in the measurements of 
the $W^{\pm}$ boson multipoles at the Tevatron collider.}. As an 
application of this method, the angular distribution of soft photons emitted 
in $\Omega^- \rightarrow \Lambda K^-$ decays has been computed in order to 
study the 
sensitivity to the anomalous magnetic moment of $\Omega^-$ \cite{martin}. 
The results obtained with this method, 
however, are not competitive with the precision attained \cite{morelos} 
using the usual spin precession technique \cite{bmt}.

  In a previous paper \cite{blg} we have analyzed the effects of 
anomalous magnetic moments of charged vector mesons ($\rho^+,\ K^{*+}$) in 
the decay rates of two-pseudoscalar radiative decays. These  
decay rates are almost insensitive to the effects of 
magnetic dipole moments unless high values are used for the infrared 
cut off photon energies. However, this reduction of photon phase space 
strongly suppresses the decay rates and make difficult their 
accurate measurement using this method.

   Following a similar approach, in this {\em Brief Report} we analyze 
the effects of $\rho^+$ and $K^{*+}$ anomalous magnetic moments in the 
decay distributions of photons emitted in the two-pseudoscalar decays of 
these vector mesons. In order to improve the sensitivity on these 
effects, we 
consider the photon energy spectrum for photons emitted at small angles 
with respect to charged pseudoscalar mesons. 

  Besides the possible experimental difficulties for reconstruction of 
these particular configurations, there are two limitations of the present 
approach. First, one should bear in mind that the decays of these 
unstable particles can not be separated from its production process as 
required in our calculations. In fact, when considering the production 
and decay mechanisms of a charged resonance in a radiative process, some 
care must be taken \cite{ginv} to maintain electromagnetic 
gauge-invariance of the amplitude in 
presence of the finite width of the resonance\footnote{ A 
corresponding analysis for the $\rho^-$ resonance in the process $\tau^- 
\rightarrow \nu_{\tau} \pi^- \pi^0 \gamma$ is underway \cite{taurho}.}.  On 
the other hand, we neglect the vector 
meson decay widths appearing in the the propagators after photon emission 
off vector mesons. While the first difficulty could be overcome by 
imposing appropriate cuts to suppress photon radiation in the production 
mechanism of the vector mesons, 
we expect that the second approximation accounts for neglecting terms of 
${\cal O}(\Gamma/M)$ as far as the photon energy is not taken as very 
low. The importance of other reasonable approximations made in  our 
calculations are discussed at the end of the paper.

 As is known, the total magnetic moment for a positively charged ($e>0$) 
vector meson of mass $M$ is given by
\begin{equation}
\mu_V = (1+\kappa) \frac{e}{2M}
\end{equation}
where $\kappa$ is the anomalous piece of the magnetic moment. In analogy 
with the $W^{\pm}$ magnetic dipole moment in the standard electroweak 
theory, $\kappa=1$ can be considered as the natural or canonical value for 
the vector mesons \cite{brodsky}. However, substantial deviations from 
this canonical value can be expected and in fact, some available 
calculations of $\Delta \kappa \equiv \kappa -1$ in the context of 
phenomenological quark models indicate values as large as $\Delta \kappa 
\sim 2.6$ \cite{hecht} for the $\rho$ meson.

  Let us start with the structure of the gauge-invariant amplitude for 
the $V^+ \rightarrow P^+ P^0 \gamma$ decay ($V^+$ is the charged vector 
meson and $P^+\ (P^0)$ is a charged (neutral) pseudoscalar):
\begin{eqnarray}
{\cal M} & = & ieg_{VPP'} \left \{ \left ( \frac{p.\epsilon^*}{p.k} - 
\frac{d.\epsilon^*}{d.k} \right ) (p-p') \cdot \eta 
+\left ( \frac{p.\epsilon^*}{p.k} - 
\frac{d.\epsilon^*}{d.k} \right ) k\cdot \eta  \right. \nonumber \\
& & \ \ \ \ \ \ \left. + \left[ 2+\frac{\Delta \kappa}{2} \left( 1 + 
\frac{\Delta^2}{M^2} \right) \right] \left(\frac{d\ \cdot 
\epsilon^*}{d \cdot k}k\cdot \eta - \epsilon^* \cdot \eta \right) \right. 
\nonumber \\
& & \ \ \ \ \ \ \left. - (2+ \Delta \kappa) \left( \frac{p\ \cdot
\epsilon^*}{p \cdot k}k\cdot \eta - \epsilon^* \cdot \eta \right) 
\frac{p\cdot k}{d \cdot k} \right\} + {\cal O}(k) \\
&=& {\cal M}_{Low} + {\cal O}(k). \nonumber
\end{eqnarray}

In the above expression, $d,\ p, \ p'$ and $k$ denote respectively the 
four-momenta of $V^+,\ P^+,\ P^0$ and the photon, $\eta \ (\epsilon^*)$ 
is the polarization four-vector of $V^+\ (\gamma)$, $\Delta^2 \equiv 
m^2_{P^+}-m^2_{P^0}$ and $g_{VPP'}$ denotes the strength of the $V^+ 
P^+P^0$ interaction. The term in curly brackets in Eq. (2) corresponds to 
the Low's amplitude \cite{low}, {\it i.e.} to the leading terms in the 
expansion of the amplitude for soft photons. The 
terms of order $k^{-1}$ arise exclusively from photon emission off the 
charges of $V^+$ and $P^+$ and the terms of order $k^0$ include also 
the photon emission amplitudes from the magnetic moment of $V^+$ 
and a contact term which is necessary for gauge invariance.

   The residual terms of order $k$ in Eq. (2) contain contributions from
the electric quadrupole moment of $V^+$ and other possible model-dependent 
pieces. 
In the following we will neglect these contributions and discuss their 
relative importance at the end of this paper.

 A straightforward calculation gives the following squared amplitude 
(with sum over vector meson polarizations): 
\begin{eqnarray}
\sum_{V^+ \ pols.} |{\cal M}_{Low} |^2 &=& e^2 g^2_{VPP'} \left \{ 
\left| \frac{p.\epsilon^*}{p.k} -\frac{d.\epsilon^*}{d.k} \right|^2 
[M^2-2\Sigma^2+\frac{\Delta^4}{M^2} ] + \frac{(\Delta \kappa)^2}{M^2} 
(p. k)^2 \left|
\frac{p.\epsilon^*}{p.k} -\frac{d.\epsilon^*}{d.k} \right|^2 \right. 
\nonumber \\
& & \ \ \ \left. -\epsilon .\epsilon^* \left[ 2 +\frac{\Delta \kappa}{2} 
(1 + \frac{\Delta^2}{M^2}) -(2+\Delta \kappa)\frac{p. k}{d. k} \right]^2 
\right\} 
\end{eqnarray}
where we have defined $\Sigma^2 \equiv m^2_{P^+}+m^2_{P^0}$.
  The previous result is in agreement with the Burnett and 
Kroll's theorem \cite{bk} (see also ref. \cite{kpz}), which establishes 
the absence of terms of ${\cal O}(k^{-1})$ in the probability for 
polarized photons. The terms of order $k^{-1}$ appears only if we consider 
the squared amplitude for polarized photons {\it and} vector mesons.

  In order to choose the decay distributions suitable to observe the 
effects of $\Delta \kappa \neq 0$, we set in the rest frame of the vector 
meson. In this case, the infrared factor in the previous result becomes:
\begin{equation}
\sum_{\gamma \ pols} 
\left| \frac{p.\epsilon^*}{p.k} -\frac{d.\epsilon^*}{d.k} \right|^2
= \frac{|\vec{p}|^2 \sin^2 \theta}{\omega^2 (E-|\vec{p}| \cos \theta)^2}
\end{equation}
where $E$ and $\omega$ are, respectively, the energies of the charged 
pseudoscalar and the photon in the rest frame of $V^+$ ($|\vec{p}| 
=\sqrt{E^2-m^2_{P^+}}$). The angle $\theta$ defines the direction of 
photon emission with respect to the charged pseudoscalar in the same frame.

 Since the $\Delta \kappa$-dependent terms in Eq. (3) start at order 
$\omega^0$, 
we can expect according to Eq. (4) that the differential decay 
distribution for photons of low energy would be more sensitive to $\Delta 
\kappa \neq 0$ if we cut the large values of $\theta$.
 Using this property, in Figures 1, 
2 and 3 we show the energy decay distributions of photons 
(normalized to the corresponding non-radiative rates, {\it i.e.} 
$(1/\Gamma_{nr})d\Gamma/dx d\cos \theta$, where $x= 2\omega/M$ ) 
in the  $\rho^+ \rightarrow \pi^+ \pi^0 
\gamma,\ K^{*+} \rightarrow K^0 \pi^+ \gamma$ and $K^{*+} \rightarrow K^+ 
\pi^0 \gamma$ decays.  The short--dashed lines in
all these plots correspond to the terms of order $\omega^{-2}$ (first
term in Eq. (3)) and arise exclusively from bremsstrahlung. The terms of 
order $\omega^0$ in Eq. (3) are plotted for three different values of the 
anomalous magnetic moment: $\Delta \kappa =-1$ (solid line), $\Delta 
\kappa =0$ (long--dashed) and $\Delta \kappa =1$ (long-short--dashed). The 
upper and lower parts in Figures 1--3 correspond respectively 
to $\theta = 10^o$ and $20^o$.  Note that the 
only unknown parameter in the plotted distributions is $\Delta \kappa$. 

  As expected, the contributions of order $\omega^0$ in Eq. (3) 
dominate over the terms of order $\omega^{-2}$ except for very low 
values of the 
photon energy. Thus, the terms dependent on $\Delta \kappa$ can be 
safely isolated by removing the ---model-independent--- contributions of 
order $\omega^{-2}$. On the other hand, while the energy distribution in 
$K^{*+} \rightarrow K^0 \pi^+ \gamma$ is largely independent of the 
precise value of $\Delta \kappa$, the best sensitivity to the effects of 
anomalous magnetic moments in $K^{*+}$ decays is observed in the $K^{*+} 
\rightarrow K^+
\pi^0 \gamma$ channel. The physical reason for this is that the 
radiation emitted by a moving charge decreases with its velocity as 
observed in Eq. (4) ($v=|\vec{p}|/E$ is smaller for $K^+$ than for 
$\pi^+$ in $K^{*+}$ decays).

   Another interesting property of these plots is the dip observed in the 
$\rho^+$ and $K^{*+} \rightarrow K^+ \pi^0 \gamma$ decays near the end of 
the photon energy spectrum. This dip (which corresponds to a vanishing 
distribution in the case $\Delta \kappa = 1$) looks similar to the null 
radiation amplitudes observed in the angular distribution of some 
radiative processes \cite{raz}. We think however that its origin is not 
the same since this dip appears for the two values of $\theta$ considered 
here and it is absent in the $K^0 \pi^+$ mode of $K^{*+}$ decay.

  Based on Eq. (3), we can give a rough estimate of the accuracy expected 
for the measurement of $\Delta \kappa$ using this method. For instance, 
if we assume $\Delta \kappa =0$ and $\theta = 15^o$, then the decay 
distributions of $\rho^+ \rightarrow \pi^+ \pi^0 \gamma$ and $K^{*+} 
\rightarrow K^+ \pi^0 \gamma$ are required to be measured with a 25 \% 
error in order to achieve an accuracy of $\delta |\Delta \kappa | =0.5$. 
This precision is almost independent of the full range of photon energies 
where the terms of order $\omega^0$ clearly dominate over the terms of 
order $\omega^{-2}$.

   Before concluding let us discuss the relative size of the 
contributions neglected in our calculation. As it was shown in ref. 
\cite{blg}, the model dependent contributions of the type $\rho^+ 
\rightarrow \pi^+ \omega \rightarrow \pi^+ \pi^0 \gamma$ and the analogous  
contributions to $K^{*+}$ decays are negligible in the decay rate. We 
can expect negligible model-dependent contributions also in the decay 
distributions considered in this paper. Indeed, model-dependent 
contributions will start at order $\omega^0$ and arise from the interference 
between model-dependent amplitudes (of order $\omega$) and the 
term of order $\omega^{-1}$, {\it i.e.} their effects will be suppressed at 
small values 
of $\theta$. Since a similar argument holds for the contribution of the 
electric quadrupole moment, we can expect that our results for the photon 
spectra would not be sizable affected when we consider photons 
emitted at small angles with respect to charged pseudoscalar mesons.

  In conclusion, the decay distributions of photons emitted in 
two-pseudoscalar radiative decays of charged vector mesons offer a valid 
alternative for a determination of their magnetic dipole moments. For 
this purpose, a measurement of the spectra of soft photons emitted at 
small angles with respect to charged pseudoscalar mesons is required. The 
model-dependent contributions and the one due to electric quadrupole 
moment of vector mesons are expected to be negligible for these 
kinematical configurations.

\

{\bf Acknowledgements.} We are grateful to A. Garc\'\i a and J. Pestieau 
for useful conversations. We acknowledge partial financial support from 
Conacyt.

\begin{figure}[t]
\leavevmode
\par
\begin{center}
\mbox{\epsfxsize=14.cm\epsfysize=16.cm\epsffile{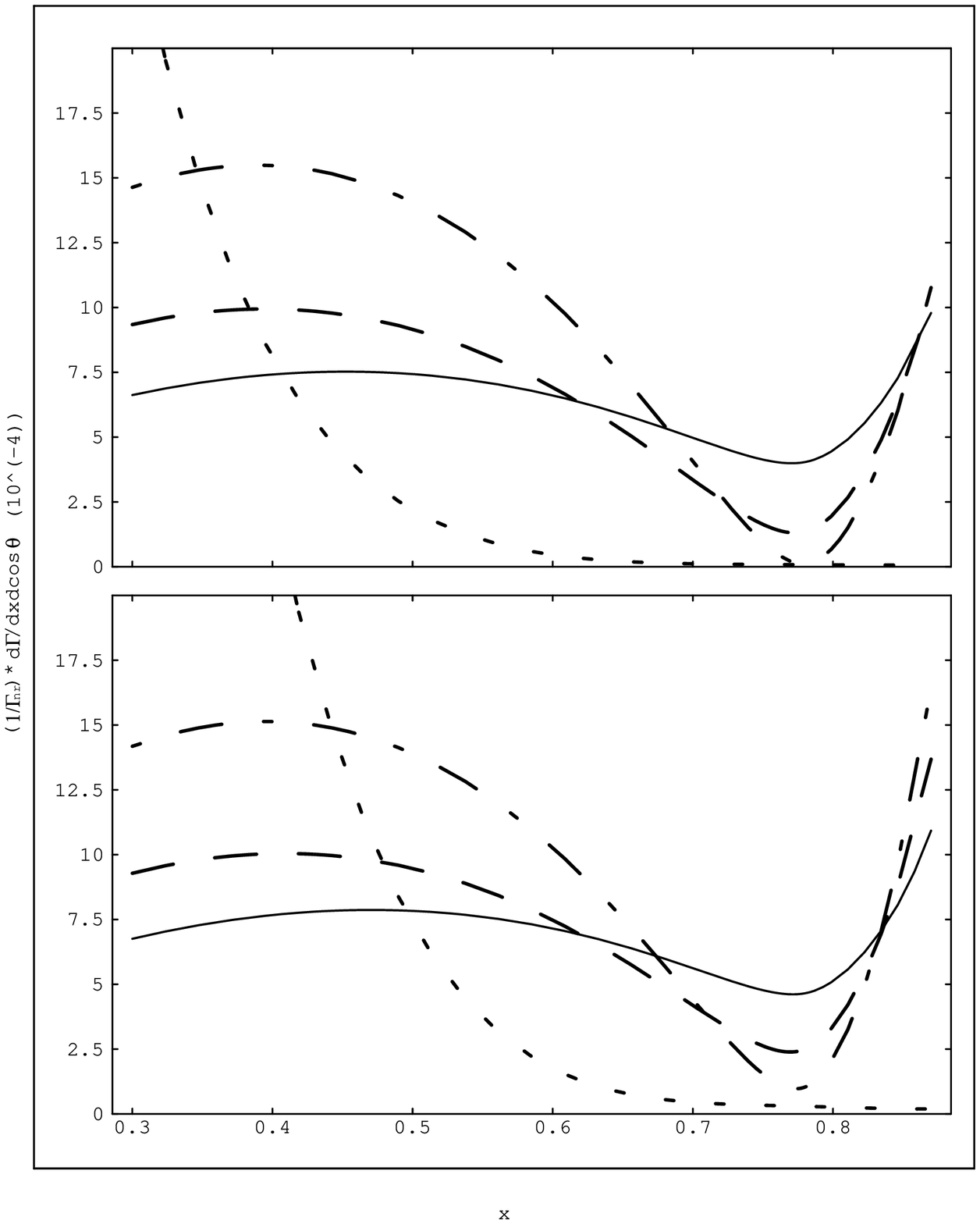}}
\end{center}
\caption{Differential decay distribution of photons in the decay $\rho^+
\rightarrow \pi^+ \pi^0 \gamma$. The
short--dashed plot corresponds to
the term of order $\omega^{-2}$ and the solid, long--dashed and
long-short--dashed plots are the terms of order $\omega^0$ when $\Delta
\kappa = -1, 0$ and $1$, respectively. The upper and lower parts  are
for $\theta =10^o$ and $20^o$, respectively.}
\end{figure}

\newpage

\begin{figure}[t]
\leavevmode
\par
\begin{center}
\mbox{\epsfxsize=14.cm\epsfysize=16.cm\epsffile{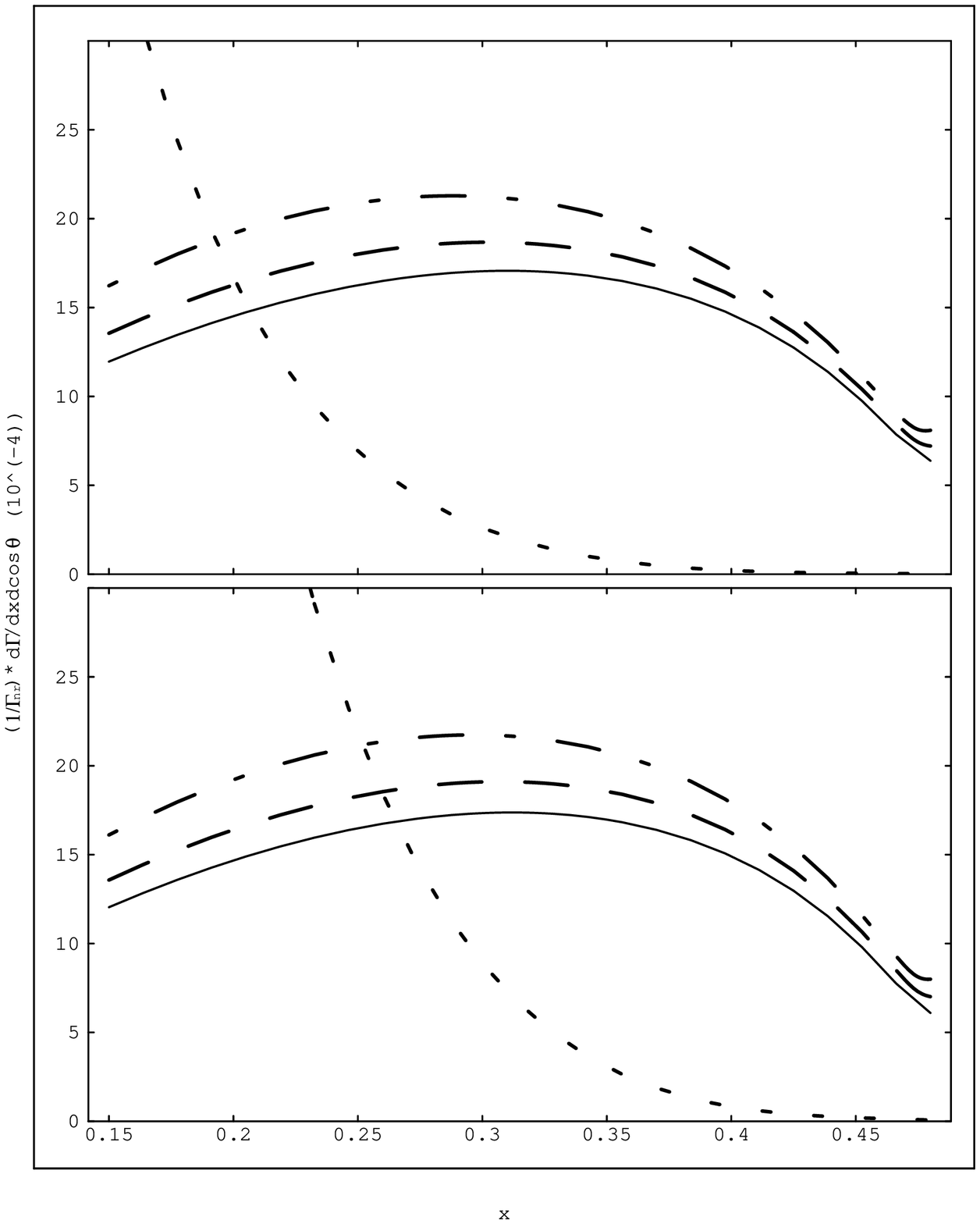}}
\end{center}
\caption{Differential decay distribution of photons in the $K^{*+}
\rightarrow K^0 \pi^+ \gamma$ decay. The
description is the same as in Figure 1.}
\end{figure}

\newpage

\

\begin{figure}[t]
\leavevmode
\par
\begin{center}
\mbox{\epsfxsize=14.cm\epsfysize=16.cm\epsffile{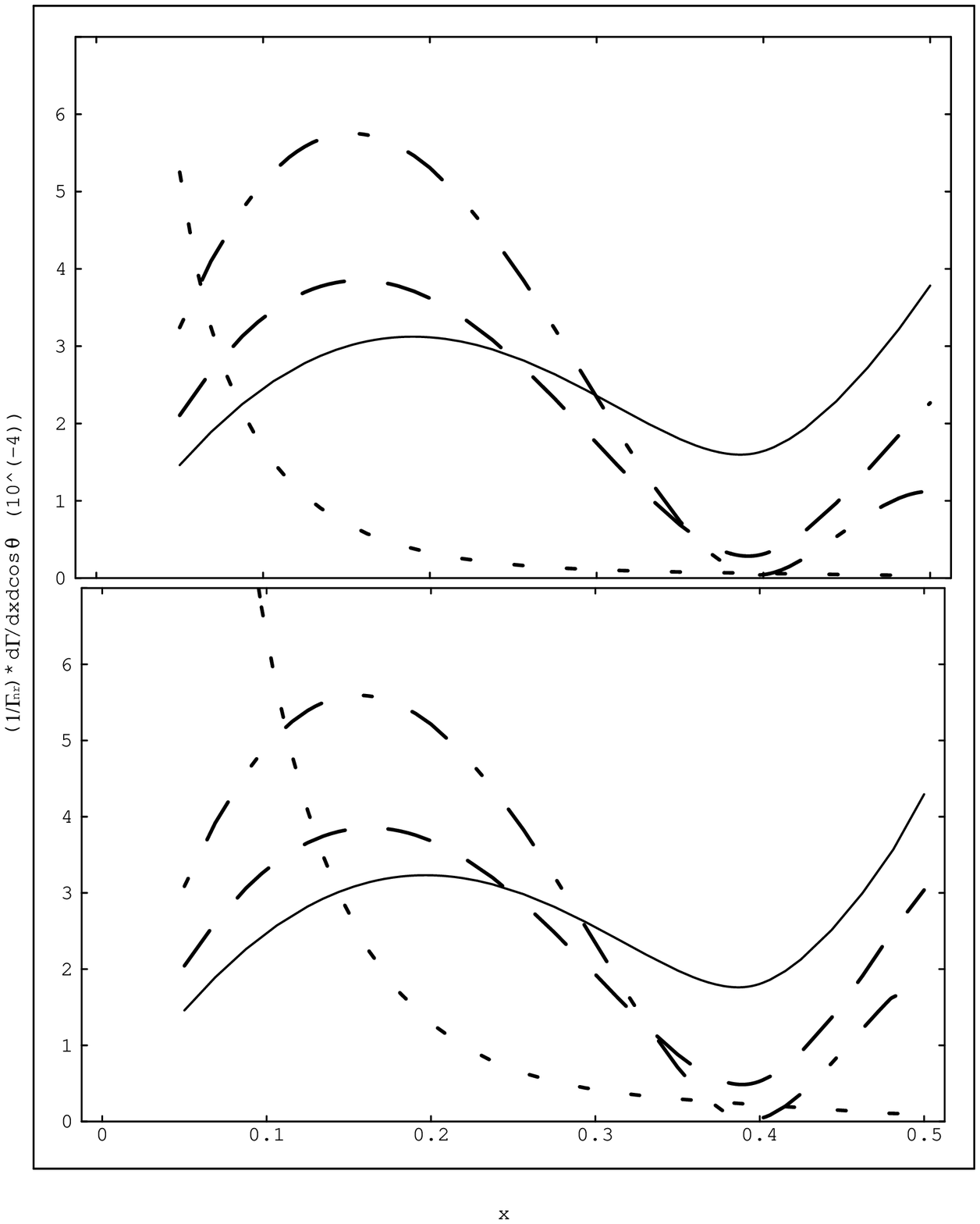}}
\end{center}
\caption{Differential decay distribution of photons in the $K^{*+}
\rightarrow K^+ \pi^0 \gamma$ decay. The description is the same as in
Figure 1.}
\end{figure}


\medskip

\begin{thebibliography}{9}
\bibitem{pdg} R. M. Barnett {\it et al.}, {\sl Review of Particle 
Physics}, Phys. Rev. {\bf D54} Part 1-I, (1996).

\bibitem{bmt} V. Bargmann, L. Michel and V. L. Teledgi, Phys. Rev. Lett. 
{\bf 2}, 433 (1959).

\bibitem{kpz} V. I. Zakharov, L. A.  Kondratyuk, L. A. Ponomarev, Sov. J. 
of Nucl. Phys. {\bf 8}, 456 (1969).

\bibitem{martin}  M. Hern\'andez, G. L\'opez Castro and J. L. Lucio M., 
Phys. Rev. {\bf D44}, 794 (1991).

\bibitem{morelos} H.T. Diehl {\it et al.}, Phys. Rev. Lett. {\bf 67}, 804 
(1991); N. B. Wallace {\it et al.}, Phys. Rev. Lett. {\bf 74}, 3732 (1995).

\bibitem{blg} A. Bramon, J. L. D\'\i az Cruz and G. L\'opez Castro, Phys. 
Rev. {\bf D47}, 5181 (1993).

\bibitem{ginv} U. Baur and D. Zeppenfeld, Phys. Rev. Lett.
{\bf 75 }, 1002 (1995);
  E.N. Argyres {\it et al.}, Phys. Lett. {\bf B358 }, 339 (1995); M. 
Beuthe, R. Gonz\'alez Felipe, G. L\'opez Castro and J. Pestieau, e-print 
hep-ph/9611434  to appear in Nucl. Phys. {\bf B} .

\bibitem{taurho} L. Florent, G. L\'opez Castro and J. Pestieau in 
preparation.

\bibitem{brodsky} 
S. J. Brodsky and J. R. Miller, Phys. Rev. {\bf D46}, 2141 (1992).

\bibitem{hecht} M. B. Hecht and B. H. J. Mc Kellar, ``Dipole moments of 
rho meson", e-preprint hep-ph/9704326 and references cited therein.

\bibitem{low}  F.E. Low, Phys. Rev. {\bf 110} (1958) 974; M. Sapir and P. 
Singer, Phys. Rev. {\bf 163}, 1756 (1967).

\bibitem{bk} T. H. Burnett and N. M. Kroll, Phys. Rev. Lett. {\bf 20}, 86 
(1968).

\bibitem{raz} R. Mikaelian, M. A. Samuel and D.Sahdev, Phys. Rev. Lett. 
{\bf 43}, 746 (1979); S. J. Brodsky and R. W. Brown, Phys. Rev. Lett. 
{\bf 49}, 966 (1982); F. Boudjema, C. Hamzaoui, M. A. Samuel  and J. 
Woodside, Phys. Rev. Lett. {\bf 63}, 1906 (1989). 

\end{thebibliography}
\end{document}